\documentclass[preprint,12pt]{elsarticle}

\usepackage{amssymb}
\usepackage{amsthm}
\usepackage{amsmath}
\usepackage{graphicx}
\usepackage{dcolumn}
\usepackage{bm}
\usepackage{hyperref}

\journal{}

\begin{document}

\begin{frontmatter}



\title{Efficiency Studies of Fast Neutron Tracking using MCNP}


\author[LANL]{Pinghan Chu}

\affiliation[LANL]{organization={Los Alamos National Laboratory},
            addressline={P.O.Box 1663}, 
            city={Los Alamos},
            postcode={87545}, 
            state={NM},
            country={USA}}

\author[LANL]{Michael R. James}
\author[LANL]{Zhehui Wang}

\begin{abstract}
Fast neutron identification and spectroscopy is of great interest to nuclear physics experiments. Using the neutron elastic scattering, the fast neutron momentum can be measured. Wang and Morris introduced the theoretical concept that the initial fast neutron momentum can be derived from up to three consecutive elastic collisions between the neutron and the target, including the information of two consecutive recoil ion tracks and the vertex position of the third collision or two consecutive elastic collisions with the timing information. Here we also include the additional possibility of measuring the deposited energies from the recoil ions. In this paper, we simulate the neutron elastic scattering using the Monte Carlo N-Particle Transport Code (MCNP) and study the corresponding neutron detection and tracking efficiency. The corresponding efficiency and the scattering distances are simulated with different target materials, especially natural silicon (92.23$\%$ $^{28}$Si, 4.67$\%$ $^{29}$Si, and 3.1$\%$ $^{30}$Si) and helium-4 ($^4$He). The timing of collision and the recoil ion energy are also investigated, which are important characters for the detector design. We also calculate the ion travelling range for different energies using the software, ``The Stopping and Range of Ions in Matter (SRIM)", showing that the ion track can be most conveniently observed in $^4$He unless sub-micron spatial resolution can be obtained in silicon.
\end{abstract}




\end{frontmatter}


\section{\label{sec:introduction}Introduction}
One important reaction for fast neutron (neutron kinetic energy {$E_{k,n}$ > $\sim$10}
 keV) detection is elastic neutron scattering by light nuclei such as $^1$H and its isotope $^2$H, $^4$He, and others. The cross sections of other reactions, {e.g.,} neutron capture, decrease rapidly with increasing neutron energy. In elastic neutron scattering, the incoming neutron can transfer a portion of its kinetic energy to the target nucleus.  Wang and Morris ~\cite{WANG:2013145} suggested the scenarios to track fast neutrons using the tracks of recoil ions and vertex positions of the collisions (method 1 and method 2 in Section~\ref{sec:methods}). Additionally, the angle between the incoming neutron and the scattered neutron may be also determined by the deposited energies from the recoil ions and the vertex positions of the two successive collisions, allowing method 3 for neutron tracking, as described in Section~\ref{sec:methods}. In this paper, we verify these three methods for fast neutron tracking using the Monte Carlo N-Particle Transport Code (MCNP)~\cite{mcnp} simulations and study the corresponding efficiencies with a cubic detector made of natural silicon (92.23$\%$ $^{28}$Si, 4.67$\%$ $^{29}$Si, and 3.1$\%$ $^{30}$Si) or $^4$He. The study here would be helpful for next-generation neutron tracking detectors, showing the limitation of the fast neutron tracking and possible designs of detectors. 

Below, we first describe the fast neutron tracking principle in Section~\ref{sec:methods}. Next, we explain the method of the MCNP Monte Carlo simulations in Section~\ref{sec:mcnp}. In Section~\ref{sec:results}, we show the simulation results, including different parameters. In Section~\ref{sec:ionrange}, we show the ion traveling range using the software of The Stopping and Range of Ions in Matter (SRIM)~\cite{Ziegler:1985} in order to evaluate the feasibility of current detector technologies. A brief summary is given at the end.
\section{Basic Idea of Fast Neutron Tracking}
\label{sec:methods}
\begin{figure}[h]
    \includegraphics[width=0.45\textwidth]{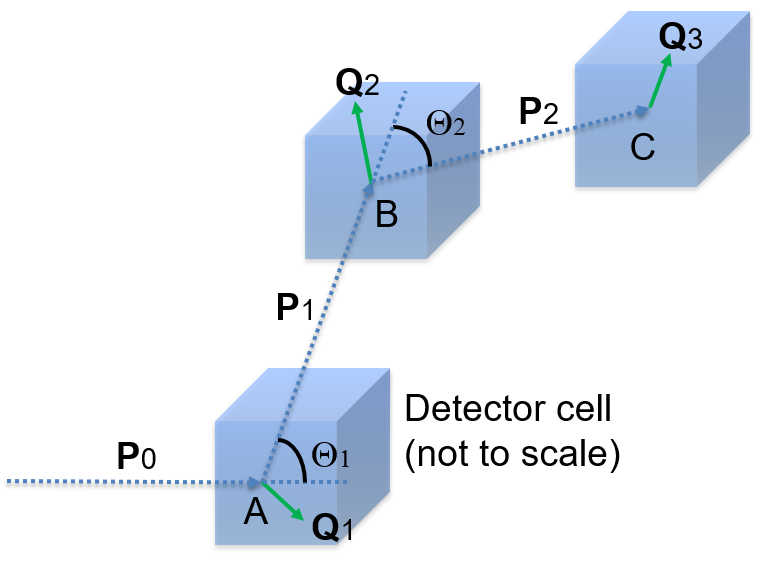}
\caption{The schematic view of three consecutive recoil ion tracks, induced by a neutron with unknown momentum $P_0$. A, B, and C are the vertices of three elastic collisions. $Q_1, Q_2$, and $Q_3$ are the momenta of the recoil ions (Adapted with permission from the author Z. Wang in Ref.~\cite{WANG:2013145}).
}
  \label{fig:threescattering}
\end{figure}

Figure~\ref{fig:threescattering} shows the schematic of three consecutive recoil ion tracks. The recoil ion momenta are $\vec{Q}_1, \vec{Q}_2$, and $\vec{Q}_3$. The initial neutron momentum is $\vec{P}_0=p_0\vec{e}_0$, where $\vec{e}_0$ is the unit vector, and the successive neutron momenta are $\vec{P}_1=p_1\vec{e}_1$ and $\vec{P}_2=p_2\vec{e}_2$, where $\vec{e}_1$ and $\vec{e}_2$ are the unit vectors which can be determined by the vertex positions A, B, and C of the collisions. The mass of a neutron is $m_n$ and the mass of the recoil ion is $m_i$.

In method 1 (M1), three ion tracks are sufficient to measure the initial momentum of the fast neutrons. Using the momentum conservation at the vertex B,
\begin{align}
    &\vec{P}_1 = \vec{Q}_2 + \vec{P}_2.
    \label{eq:p1q2p2}
\end{align}

Dotting with $\vec{e}_1$ or $\vec{e}_2$, Equation~\eqref{eq:p1q2p2} becomes
\begin{align}
    p_1 = \vec{Q}_2\cdot\vec{e}_1 + p_2 \vec{e}_2\cdot\vec{e}_1
\end{align}
or
\begin{align}
    p_1 \vec{e}_1\cdot\vec{e}_2= \vec{Q}_2\cdot\vec{e}_2 + p_2,
\end{align}
respectively. Therefore,
\begin{align}
    p_1 =& \vec{Q}_2\cdot\vec{e}_1 + (p_1\vec{e}_1\cdot\vec{e}_2-\vec{Q}_2\cdot\vec{e}_2)\vec{e}_2\cdot\vec{e}_1\notag\\
     \rightarrow p_1=& \frac{\vec{Q}_2\cdot\vec{e}_1-(\vec{Q}_2\cdot\vec{e}_2)\cos{\Theta_2}}{1-\cos^2{\Theta_2}}
\end{align}
and
\begin{align}
    p_2 =& p_1\vec{e}_1\cdot\vec{e}_2-\vec{Q}_2\cdot\vec{e}_2\notag\\
    =& \frac{\vec{Q}_2\cdot\vec{e}_1-(\vec{Q}_2\cdot\vec{e}_2)(\vec{e}_2\cdot\vec{e}_1)}{1-(\vec{e}_1\cdot\vec{e}_2)^2}(\vec{e}_1\cdot\vec{e}_2)-\vec{Q}_2\cdot\vec{e}_2\notag\\
    =& -\frac{\vec{Q}_2\cdot\vec{e}_2-(\vec{Q}_2\cdot\vec{e}_1)\cos{\Theta_2}}{1-\cos^2{\Theta_2}}
\end{align}
where $\vec{e}_1\cdot\vec{e}_2=\cos{\Theta_2}$.

Using the momentum conservation at the vertex A, the initial linear momentum of the fast neutron would be
\begin{align}
    \vec{P}_0&=\vec{Q}_1+\vec{P}_1 \notag\\
    &=\vec{Q}_1+\frac{\vec{Q}_2\cdot\vec{e}_1-(\vec{Q}_2\cdot\vec{e}_2)\cos{\Theta_2}}{1-\cos^2{\Theta_2}}\vec{e}_1.
    \label{eq:p0_M1}
\end{align}

Here, we show that the vectors $\vec{Q}_1, \vec{Q}_2$, and $\vec{e}_1, \vec{e}_2$ (or the vertices A, B, C) are sufficient to determine $\vec{P}_0$. 

In method 2 (M2), the time-of-flight (TOF) of the fast neutron can be determined between ion tracks. If the time delay between the vertex A and B is measured as $\tau_1$, the velocity of the fast neutron is 
\begin{align}
    v_1 = \frac{\overline{\text{AB}}}{\tau_1}
\end{align}
where $\overline{\text{AB}}$ is the distance between vertex A and B, and the momentum of the fast neutron is
\begin{align}
    p_1 = \gamma m_n v_1\label{eq:p1}
\end{align}
where $\gamma = 1/\sqrt{1-(v_1/c)^2}$. Therefore, the initial linear momentum of the fast neutron, 
\begin{align}
\vec{P}_0 &= \vec{Q}_1 + p_1\vec{e}_1 \notag\\
&=\vec{Q}_1 + \gamma m_n v_1 \vec{e}_1,
\label{eq:p0_M2}
\end{align}
can be determined by $\vec{Q}_1$, $\vec{e}_1$ and $\tau_1$ and there is no need of the third collision vertex C.

For solid detectors, it is challenging to measure the recoil ion tracks. Instead, it is straightforward to measure the recoil ion energy. Here, we apply method 3 (M3), where we can measure the vertex A and B and the corresponding time $\tau_1$ between two vertices. As Equation~\eqref{eq:p1}, $v_1=\overline{\text{AB}}/\tau_1$, $p_1=\gamma m_n v_1$ and then $\vec{P_1} = p_1 \vec{e}_1$. We can also measure the kinetic energy of the recoil ion, i.e., the deposit energy, $E_{k,Q_1}$, at the vertex A, so that the total energy, $E_{Q_1}$, of the recoil ion at the vertex A is
\begin{align}
    E_{Q_1}^2 = (Q_1 c)^2+(m_i c^2)^2 = (E_{k,Q_1}+m_ic^2)^2\label{eq:Q1}
\end{align}
where $m_i$ is the mass of the recoil ion.

Using the energy and momentum conservation, we have  
\begin{align}
    E_{0}+m_ic^2=&E_{1}+E_{Q_1},\label{eq:E0}\\
    \vec{P}_{0}=&\vec{Q}_1+\vec{P}_{1}\label{eq:p0}
\end{align}
where $E_{0}$ is the total energy of the incoming neutron and $E_{1}$ is the total energy of the scattered neutron. 

$p_0$ can be derived from Equation~\eqref{eq:E0} using
\begin{align}
    &\sqrt{(p_0c)^2+(m_nc^2)^2}+m_ic^2 \nonumber\\=& \sqrt{(p_1 c)^2 + (m_n c^2)^2} + \sqrt{(Q_1 c)^2 + (m_i c^2)^2}\label{eq:Q14}
\end{align}
where $p_1$ is defined in Equation~\eqref{eq:p1} and $Q_1$ is defined in Equation~\eqref{eq:Q1} .

Equation~\eqref{eq:p0} becomes
\begin{align}
    Q_1^2 =& p_0^2+p_1^2-2\vec{P}_0\cdot\vec{P}_1\nonumber\\
    =& p_0^2+p_1^2-2p_0 p_1\cos{\Theta_{1}}.\label{eq:Q15}
\end{align}

Therefore, $\Theta_1$ can be derived from Equation~\eqref{eq:Q15} as
\begin{align}
    \cos{\Theta_1} = \frac{p_0^2+p_1^2-Q_1^2}{2p_0p_1}
\end{align}
where $p_0$ in Equation~\eqref{eq:Q14}, $p_1=\gamma m_n v_1$ and $E_{k,Q_1}$ are measurable parameters.

\section{MCNP Simulations}
\label{sec:mcnp}
In order to evaluate the methods tracking fast neutrons using consecutive recoil ion tracks~\cite{WANG:2013145}, we apply the Monte Carlo N-Particle Transport Code (MCNP)-6.2~\cite{mcnp} developed by Los Alamos National Laboratory (LANL) to simulate the fast neutrons scattering in natural silicon (Si) of the density $\rho_{\text{Si}} = 2.329~\text{g}/\text{cm}^{3}$ and helium-4 ($^4$He). 

\begin{figure}[h]

    \includegraphics[width=0.98\textwidth]{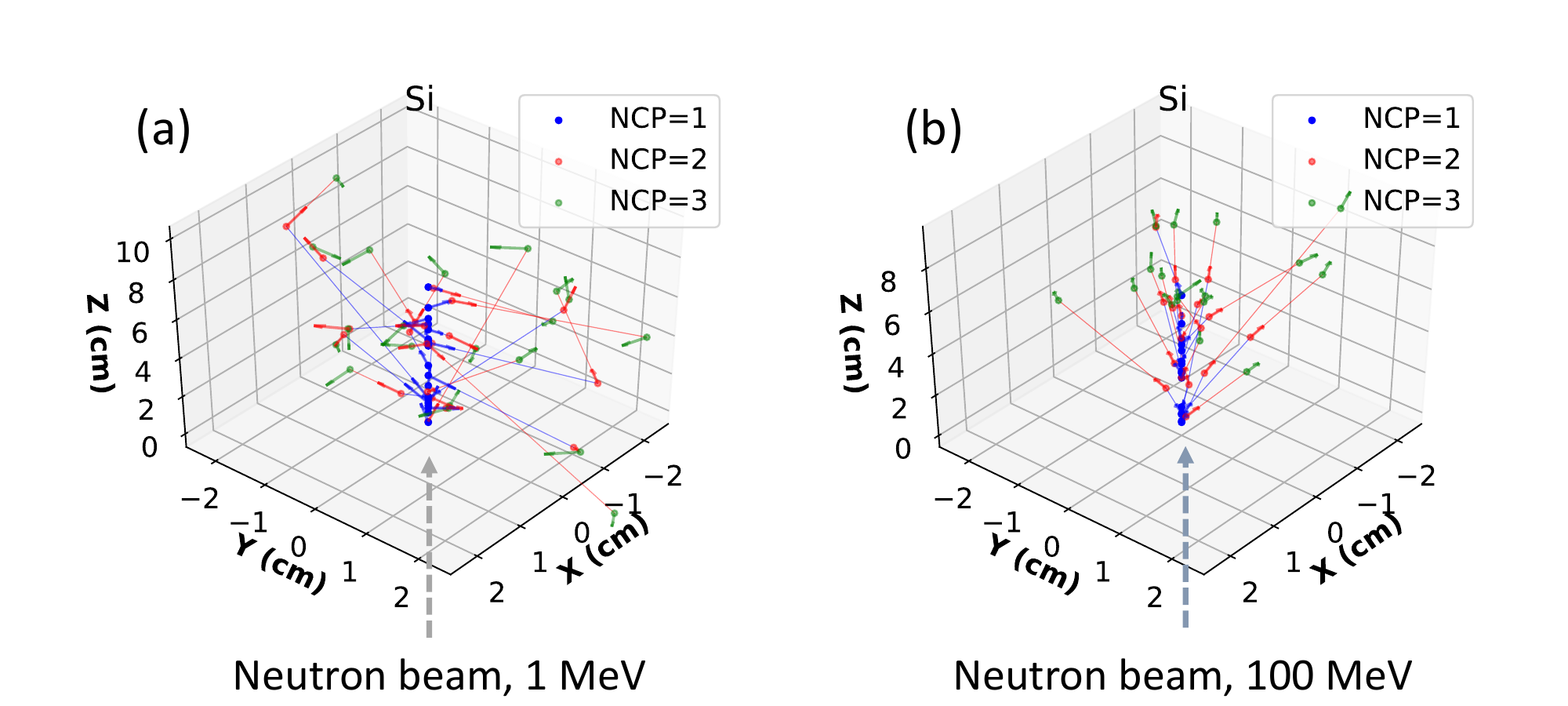}    
    \includegraphics[width=0.98\textwidth]{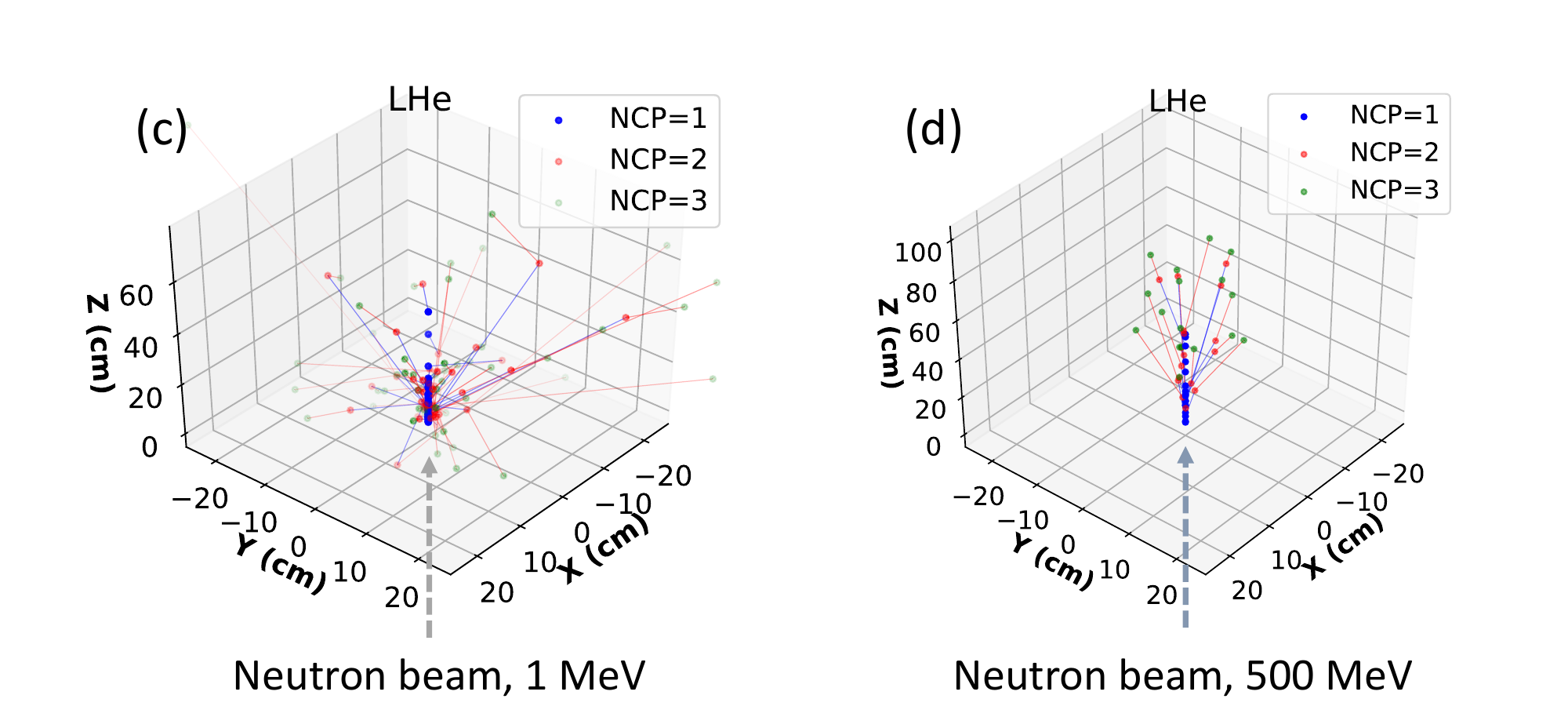}  
\caption{The scattering plots. ``NCP'' means the number of collisions in history. Blue points are the vertex A (the first collision), red points are the vertex B (the second collision), and green points are the vertex C (the third collision). The arrows are the velocity direction. (\textbf{a}) 1 MeV neutron beam with the silicon target. (\textbf{b}) 100 MeV neutron beam with the silicon target. (\textbf{c}) 1 MeV neutron beam with the liquid helium target. (\textbf{d}) 500 MeV neutron beam with the liquid helium target.}
  \label{fig:scattering}
\end{figure}

In the simulation, we consider a point-like mono-energetic neutron beam shooting into a cubic target, i.e., a 10 cm$^3$ Si cube or a 1 m$^3$ He cube. The silicon target is natural silicon while the $^4$He target has the form of liquid helium (LHe) ($\rho_{\text{LHe}} = 0.125~\text{g}/\text{cm}^{3}$), gas helium (GHe) of 50 bar ($\rho_{\text{GHe,50bar}} = 7.749\times 10^{-3} \text{g}/\text{cm}^{3}$) and 10 bar ($\rho_{\text{GHe,10bar}} = 1.638\times 10^{-3} \text{g}/\text{cm}^{3}$) in the same volume at the room temperature, 18 $^{\circ}$C. For each energy, from 100 keV to 1000 MeV, one million neutrons are generated in order to study the efficiency, the corresponding scattering ranges, and other properties. Figure~\ref{fig:scattering} shows a few corresponding scattering events using 1 MeV neutrons and 100 MeV neutrons with the Si target and 1 MeV neutrons and 500 MeV neutrons with the LHe target. Later, we show the scatter is more isotropic for low-energy neutrons and anisotropic for high-energy neutrons. This is intrinsic in the scattering theory, depending on the mass of the nucleus. 

MCNP allows users to record ``particle track output'' (PTRAC). PTRAC files collect the information of each collision/reaction, including reaction type, particle type, track number, coordinate and time of collision, and velocity direction and kinetic energy of the emitting particle. Here, we consider events of three consecutive elastic recoils for M1 and two consecutive recoils for M2 and M3. There is no ion track information in PTRAC. We calculate this information using the incoming particle and the scattered particle. The vectors $\vec{e}_1$ and $\vec{e}_2$ are derived from the collision vertex A, B, and C. The vectors $\vec{Q}_1$ and $\vec{Q}_2$ are derived using the energy and momentum conservation at the vertex A and B. For example, at the vertex B, the energy of the incoming particle is 
$E_1 = E_{k,1}+m_n c^2$, where $E_{k,1}$ is the kinetic energy recorded in PTRAC. Similarly, the energy of the scattered particle is $E_{2} = E_{k,2}+m_n c^2$. Therefore, $p_1c = \sqrt{E_1^2-(m_n c^2)^2}$ and $p_2c = \sqrt{E_2^2-(m_n c^2)^2}$. Since PTRAC records the velocity direction, we can derive $\vec{P}_1$ and $\vec{P}_2$. Then, the momentum vector components can be derived as $\vec{Q}_2 = \vec{P}_1-\vec{P}_2$. Using $\vec{e}_1$ and $\vec{e}_2$ from the vertex A, B, and C, and $\vec{Q}_1$ and $\vec{Q}_2$ from the momentum conservation, we can derive the information in Equation~\eqref{eq:p0_M1}. Using $\vec{e}_1$ from the vertex A and B, $\vec{Q}_1$ from the momentum conservation, and the time $\tau_{1}$ in PTRAC output, we can derive the information in Equation~\eqref{eq:p0_M2}.

\section{Results and Discussion}
\label{sec:results}

\begin{figure}[h]
    \includegraphics[width=0.49\textwidth]{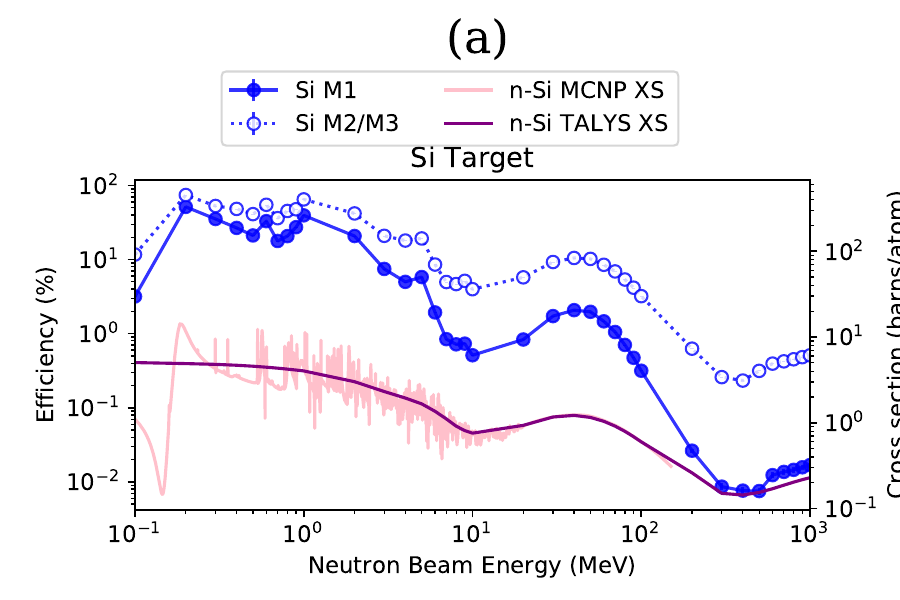}
    \includegraphics[width=0.49\textwidth]{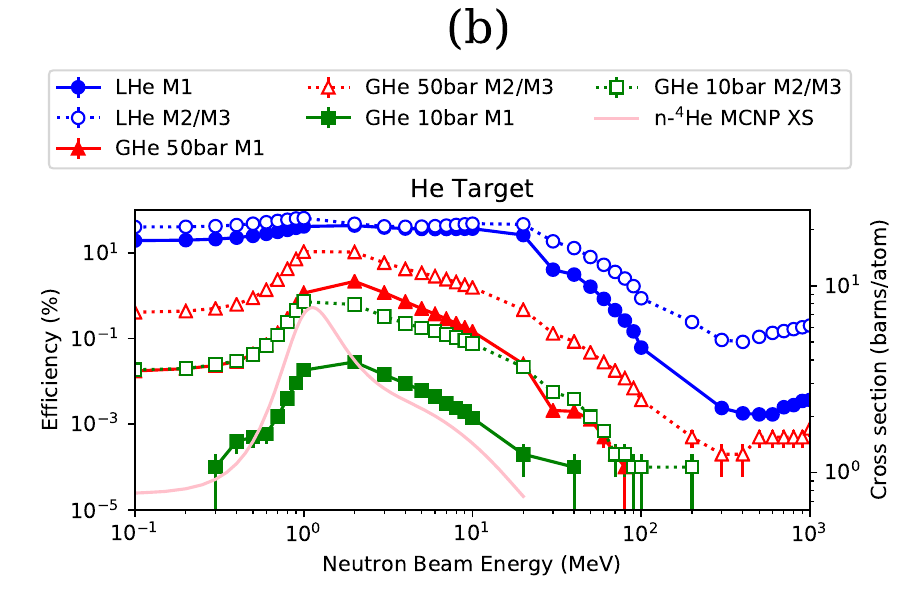}
\caption{(\textbf{a}) Efficiency with a $10$ cm$^3$ Si cube. Blue solid curve shows the triple consecutive scattering efficiency using the M1 while the blue dotted curve shows the two consecutive scattering efficiency using the M2. The pink curve shows the neutron--silicon cross section used in the MCNP-6.2 and the purple curve shows the cross section calculated by TALYS-1.95. (\textbf{b}) Efficiency with a \mbox{$1$ m$^3$} He cube. Blue curve is liquid, red is 50 bar gas, and green is 10 bar gas, while the solid curves represent the one using the M1 and the dotted curves represent the one using the M2. The pink curve shows the neutron-$^4$He cross section used in the MCNP, while TALYS-1.95~\cite{talys:1.9}  
does not calculate the neutron-$^4$He \mbox{cross section.}}
  \label{fig:efficiency}
\end{figure}

Figure~\ref{fig:efficiency} shows the efficiency and the elastic cross section of the neutron and the target as a function of the input neutron energy of M1, M2, and M3 in the Si and He target. Here the efficiency is defined as the ratio of simulation events passing the selection criteria (two/three consecutive elastic recoils) to the total number of input neutrons. The MCNP cross section is used in the MCNP simulation, while the TALYS-1.95~\cite{Koning:2007,talys:1.9} cross section is just for the comparison. The cross section of neutron--silicon has two resonances around \mbox{10 MeV} and 300 MeV.  

Since the M1 requires three consecutive scatterings while the M2/M3 requires two consecutive scatterings with the timing information, the efficiency of M2/M3 is always higher than M1. In the silicon target, the M1 efficiency is above 10\% for the energy below 10 MeV and significantly drops to <$\sim$1\% for the neutron energy larger than 10 MeV. The M2/M3 efficiency is a few factors larger than the M1 efficiency for the energy below 10 MeV and about one order larger than the M1 efficiency for the energy above 10 MeV. 

The efficiency also significantly depends on the target density. In the LHe target, the efficiency is larger than 10\% for the energy below 20 MeV and significantly drops for the energy above 20 MeV. The M2 efficiency is slightly larger than the M1 efficiency for the energy lower than 20 MeV but significantly larger for the energy above 20 MeV. The efficiency of the gas helium target of 50 bar is at least 1 order smaller than the liquid helium target, and the efficiency of the gas helium of 10 bar is much smaller. The pattern of the efficiency is consistent with the cross section used in the MCNP and calculated by TALYS.

\begin{figure}[h]
    \includegraphics[width=0.49\textwidth]{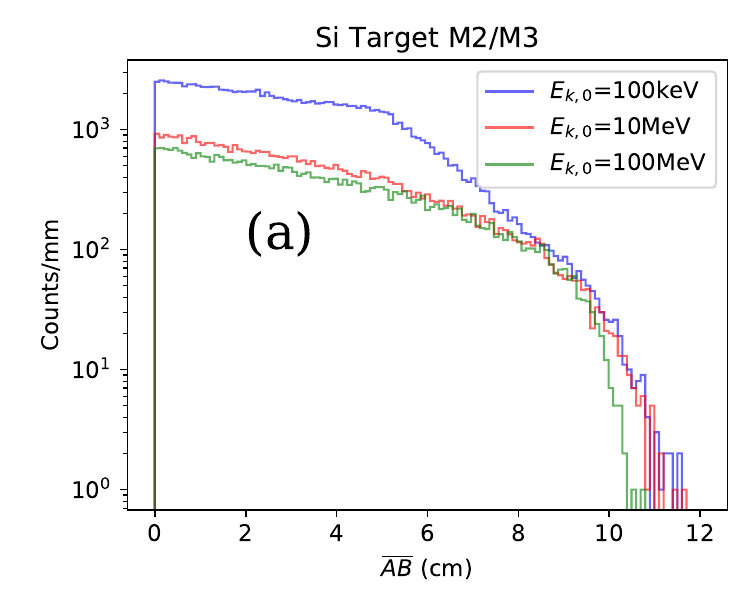}
    \includegraphics[width=0.49\textwidth]{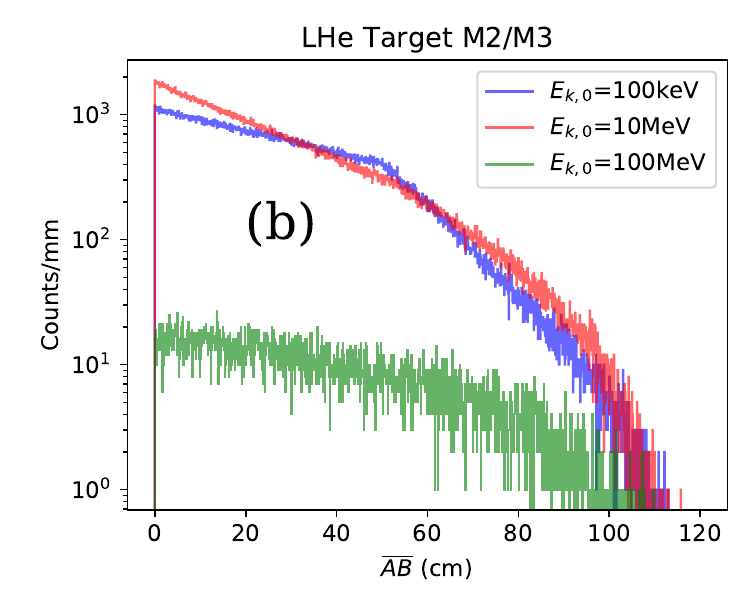}
\caption{The distribution $\overline{\text{AB}}$ for the M3 in a (\textbf{a}) 10 cm$^3$ Si cube and (\textbf{b}) 1 m$^3$ LHe cube. It shows that a 10 cm Si cube can cover most scattering events.}
  \label{fig:AB_distribution}
\end{figure}

\begin{figure}[h]
    \includegraphics[width=0.49\textwidth]{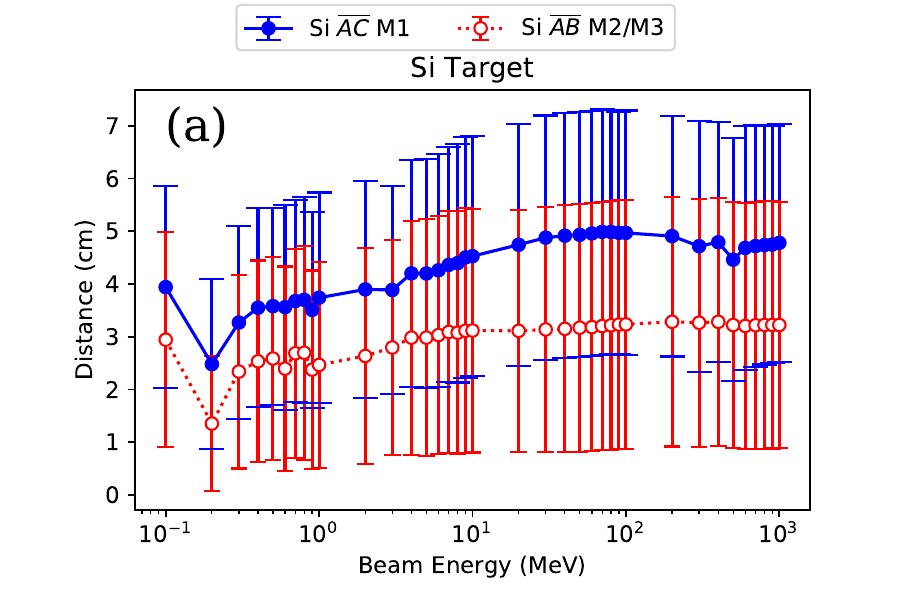}
    \includegraphics[width=0.49\textwidth]{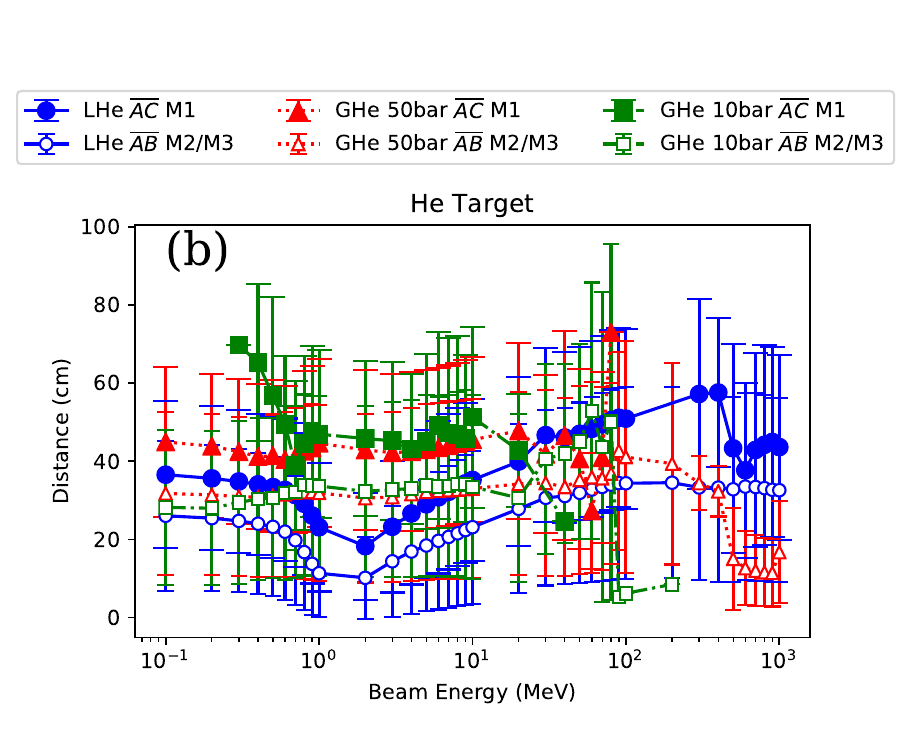}
\caption{The mean of the distance of $\overline{\text{AC}}$ (solid line) for the M1 and $\overline{\text{AB}}$ (dashed line) for the M2. \mbox{(\textbf{a}) In} a 10 cm$^3$ Si cube, blue curve shows the case for the M1. Red curve shows the case for the M2. \mbox{(\textbf{b}) In} a 1 m$^3$ He cube, blue curve shows the case for the liquid helium, red curve shows the case for the gas helium of 50 bar, and green curve shows the case for the gas helium of 10 bar.}
  \label{fig:AC_AB}
\end{figure}

Figure~\ref{fig:AB_distribution} shows that most scattering events can be detected by the 10 cm Si cube. Figure~\ref{fig:AC_AB}(a) shows that the mean distances of $\overline{\text{AC}}$ and $\overline{\text{AB}}$ in a silicon cube are about 4 to 6 cm and 2 to 3 cm, respectively, which is a reasonable value considering a 10 cm$^3$ cube. The scattering scale in a Si target is consistent within 10 cm, which is about the size of the silicon pixel detectors that we can construct. Figure~\ref{fig:AC_AB}(b) shows the mean distance of $\overline{\text{AC}}$ and $\overline{\text{AB}}$ in a $^4$He cube in different states, including liquid, gas of 50 bar, and gas of \mbox{10 bar}. The scattering scale in a He target is consistent within 1 m, which is about the size of common liquid helium Dewars.  

\begin{figure}[h]
    \includegraphics[width=0.49\textwidth]{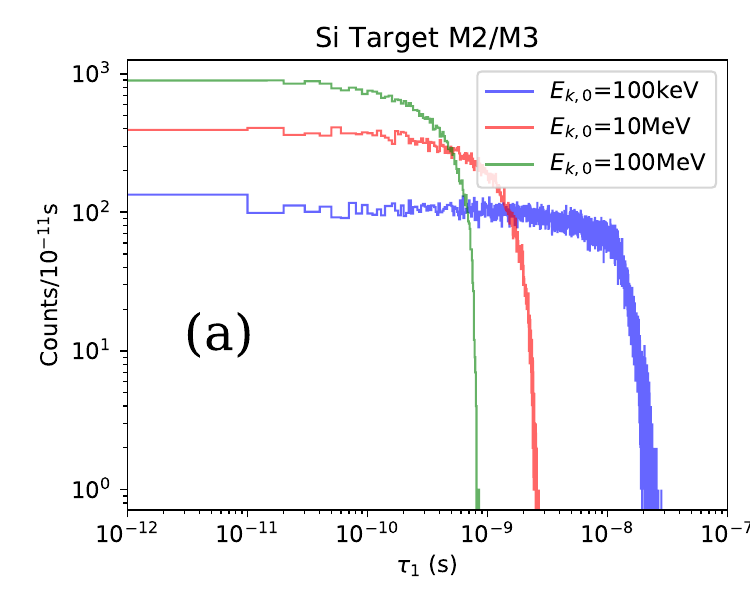}
    \includegraphics[width=0.49\textwidth]{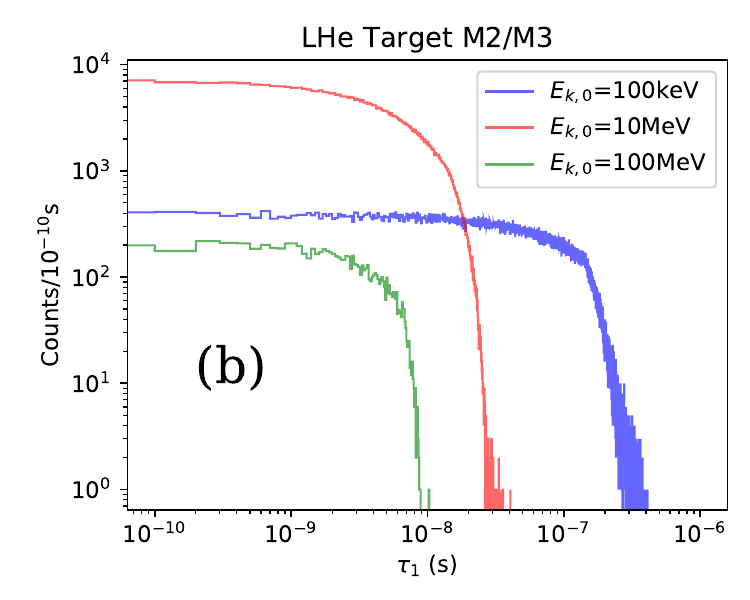}
\caption{The distribution of the time between $\overline{\text{AB}}$, $\tau_{1}$ for (\textbf{a}) silicon target; (\textbf{b}) liquid helium target with the M3. }
  \label{fig:TimeAB_distribution}
\end{figure}

\begin{figure}[h]
    \includegraphics[width=0.49\textwidth]{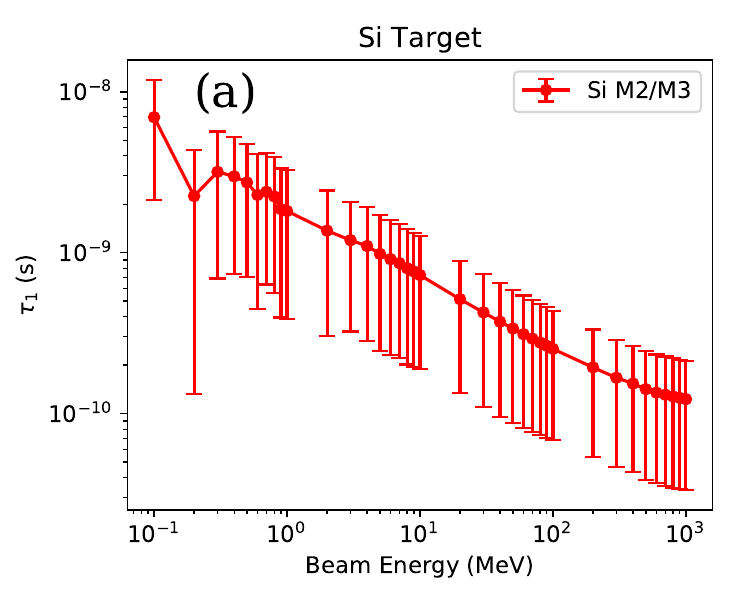}
    \includegraphics[width=0.49\textwidth]{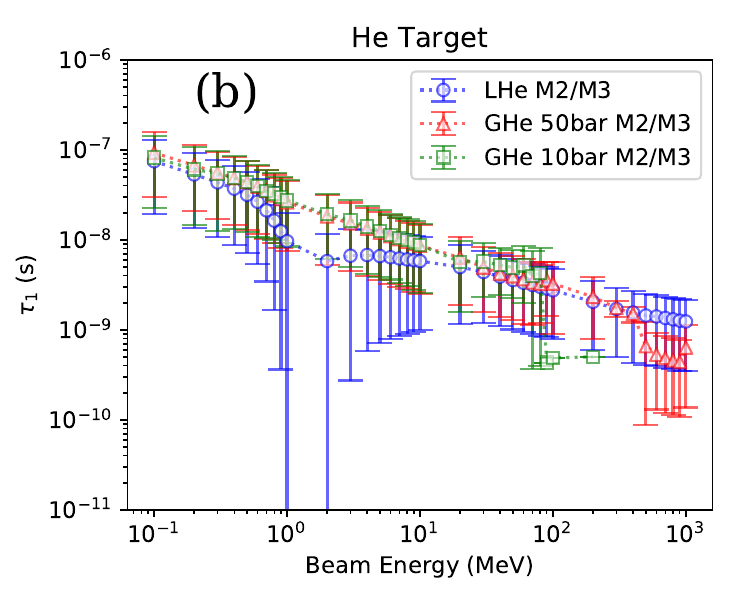}
\caption{The time between $\overline{\text{AB}}$, $\tau_{1}$ for (\textbf{a}) silicon target; (\textbf{b}) $^4$He target with the M2. }
  \label{fig:TimeAB}
\end{figure}

Figure~\ref{fig:TimeAB_distribution} shows the distribution of the time $\tau_1$. Figure~\ref{fig:TimeAB} shows the average time $\tau_{1}$. The time between the vertex A and B depends on the neutron velocity and the distance between A and B. Since the distance is roughly a constant, as shown in Figure~\ref{fig:AC_AB}, the time is roughly linearly dependent on the beam energy in the log scale. In order to measure $\tau_1$, it is necessary to have at least a time resolution of about $10^{-10}$ s for the silicon target and \mbox{$10^{-9}$ s} for the helium target.

\begin{figure}[h]
\vspace{-0.1cm}
  \hspace*{-0.3cm}
    \includegraphics[width=0.49\textwidth]{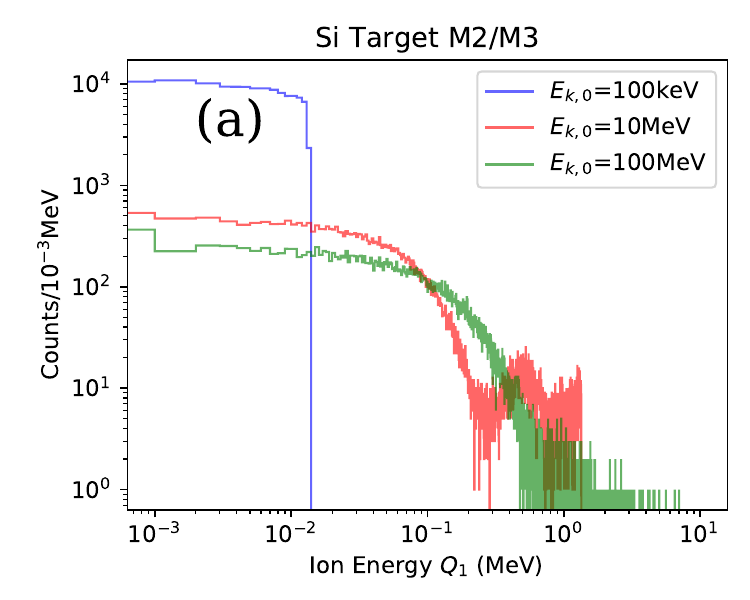}
    \includegraphics[width=0.49\textwidth]{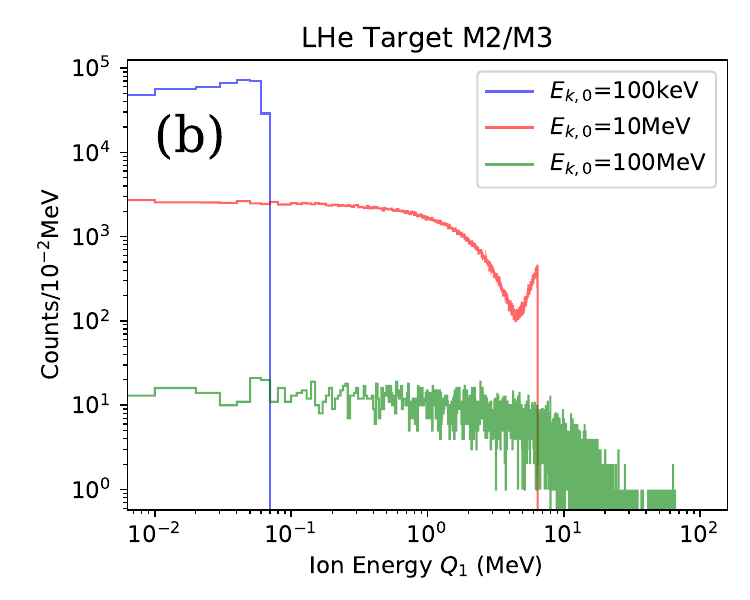}
\caption{The energy distribution of the ion with the momentum $Q_1$ in (\textbf{a})  the silicon target and \mbox{(\textbf{b}) in} the liquid helium target.}
  \label{fig:Q1Q2_ionerg_distribution}
\end{figure}

\begin{figure}[h]
  \hspace*{-0.3cm}
    \includegraphics[width=0.49\textwidth]{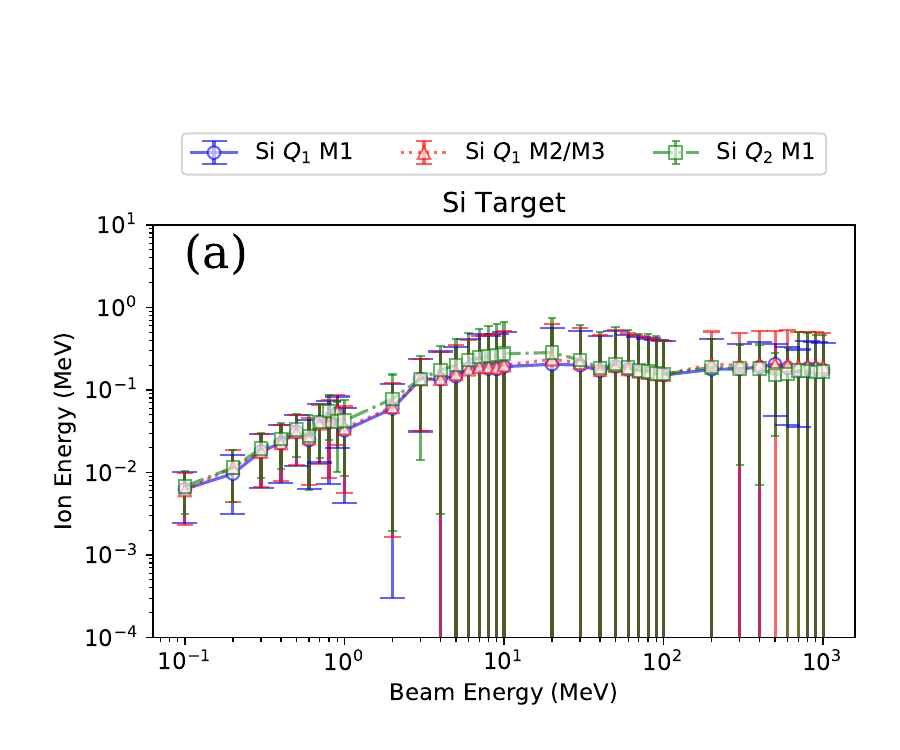}
    \includegraphics[width=0.49\textwidth]{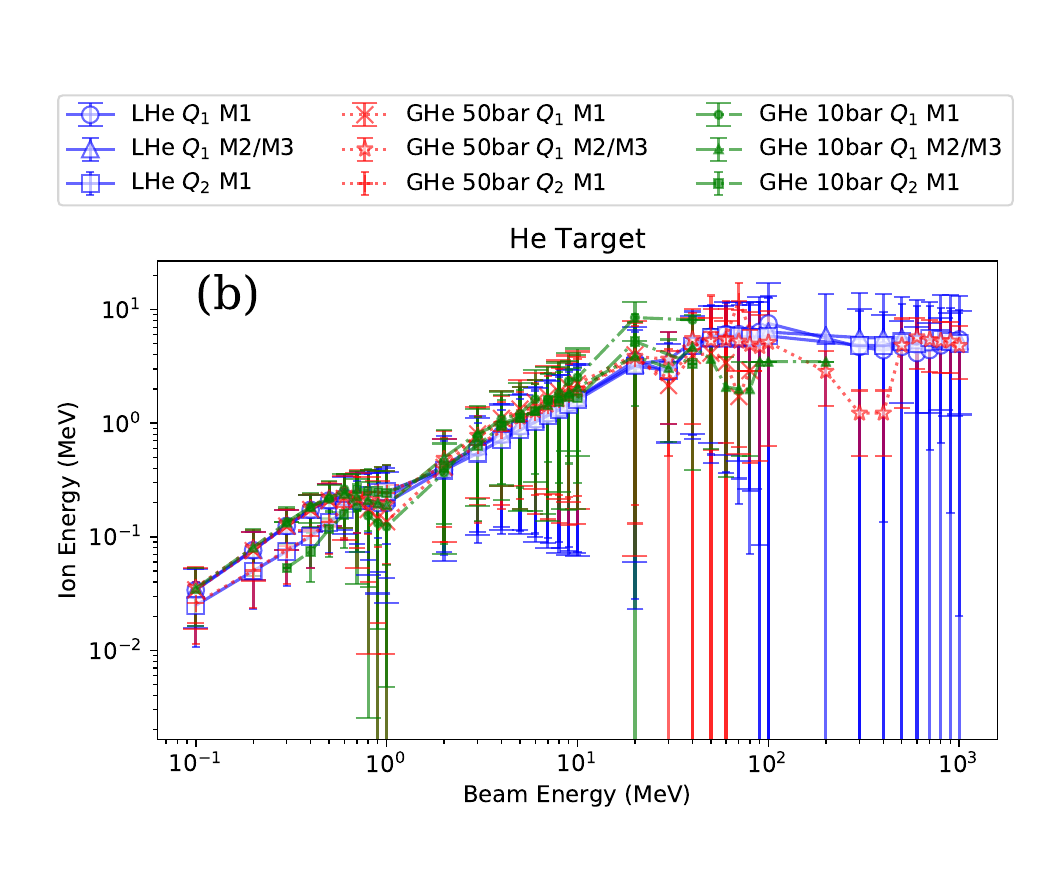}
\caption{The energy of the ion with the momentum $Q_1$ and $Q_2$. (\textbf{a}) The energy scale is about 0.2 MeV and below in the silicon target. (\textbf{b}) The energy scale is about 10 MeV and below in the helium target.}
  \label{fig:Q1Q2_ionerg}
\end{figure}

Figure~\ref{fig:Q1Q2_ionerg_distribution} shows the distribution of the ion energy with the momentum $\vec{Q}_1$ and $\vec{Q}_2$. Figure~\ref{fig:Q1Q2_ionerg} shows the average ion energy with the momentum $\vec{Q}_1$ and $\vec{Q}_2$, which is equal to the neutron deposit energy. Neutrons can deposit more energy in the elastic scattering in the $^4$He target than the silicon target because of the lighter mass of $^4$He. The deposit energy scale due to the elastic scattering is about 0.2 MeV and below in the silicon target and about 10 MeV and below in the $^4$He target. Higher deposited energy can induce nuclear reactions other than the elastic scattering, e.g., inelastic scattering.


\begin{figure}[h]
    \includegraphics[width=0.49\textwidth]{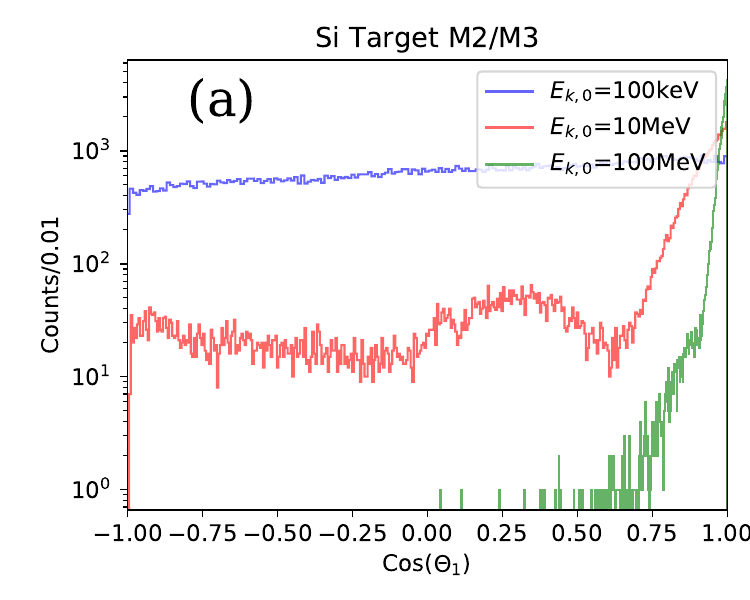}
    \includegraphics[width=0.49\textwidth]{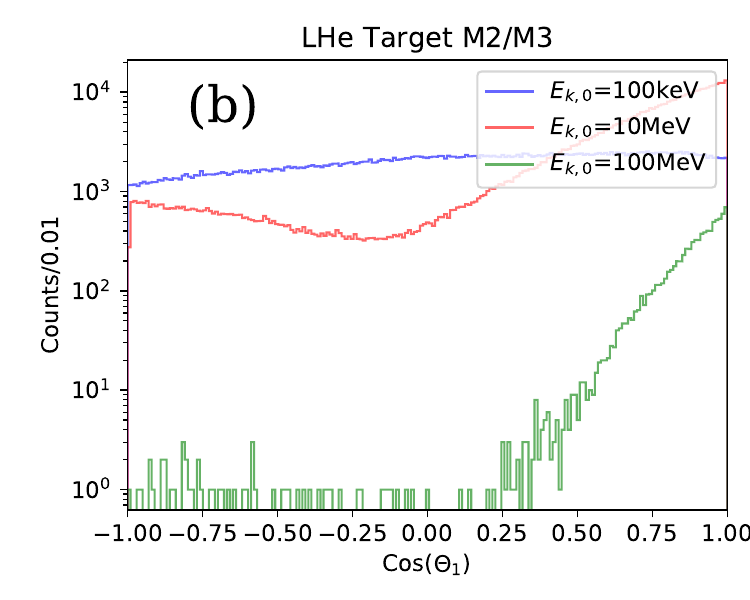}
\caption{The distribution of $\cos(\Theta_1)$ for (\textbf{a}) silicon target, (\textbf{b}) liquid helium target. }
  \label{fig:costheta_distribution}
\end{figure}

\begin{figure}[h]
    \includegraphics[width=0.49\textwidth]{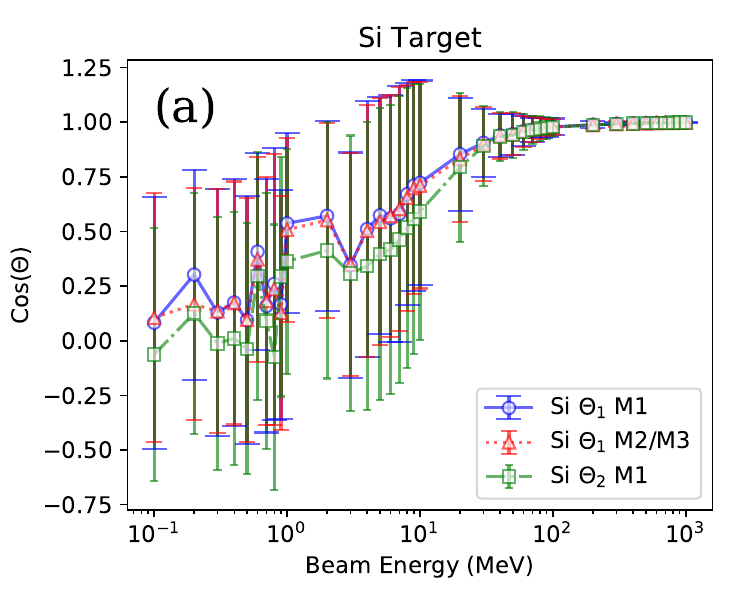}
    \includegraphics[width=0.49\textwidth]{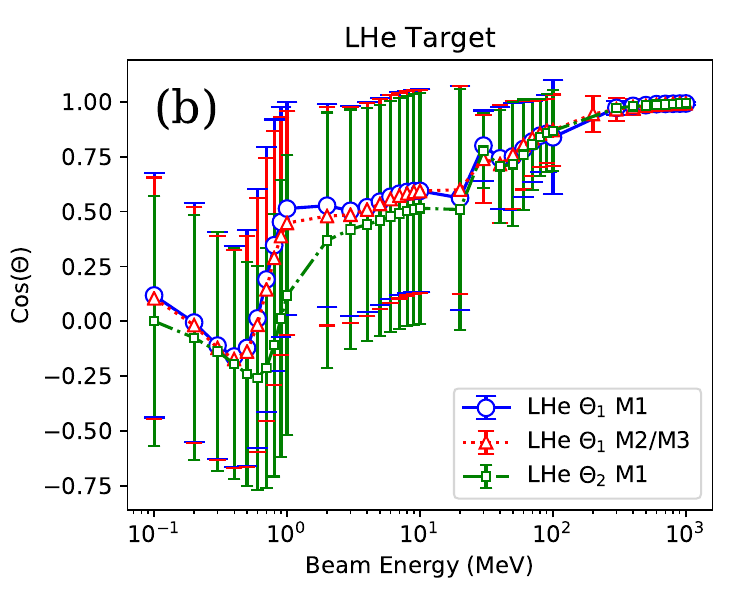}
\caption{The $\cos(\Theta)$ for $\Theta_1$ and $\Theta_2$. (\textbf{a}) 
 silicon target, (\textbf{b}) liquid helium target. }
  \label{fig:costheta}
\end{figure}

Figure~\ref{fig:costheta_distribution} shows the distribution of the  $\cos{\Theta_1}$.  Figure~\ref{fig:costheta} shows the $\cos{\Theta}$ of $\Theta_1$ and $\Theta_2$. While the beam energy is below 1 MeV, the mean of $\cos{\Theta}$ is close to zero, i.e., the elastic scattering can be any angle. When the beam energy increases above 1 MeV, the mean of $\cos{\Theta}$ is positive, i.e., the angle is less than 90$^{\circ}$ and the scattered particle tends to move forward. For the beam energy larger than 100 MeV, $\cos{\Theta}\sim 1$, which means that the scattering angles are extremely small and the scattered particles just fly forward. Any reflection will deposit enormous energy and generate nuclear reactions other than elastic scattering. Figure~\ref{fig:scattering} also shows the same pattern as isotropic scatter for low-energy neutrons and forward scatter for high-energy neutrons.

\begin{figure}[h]
    \includegraphics[width=0.49\textwidth]{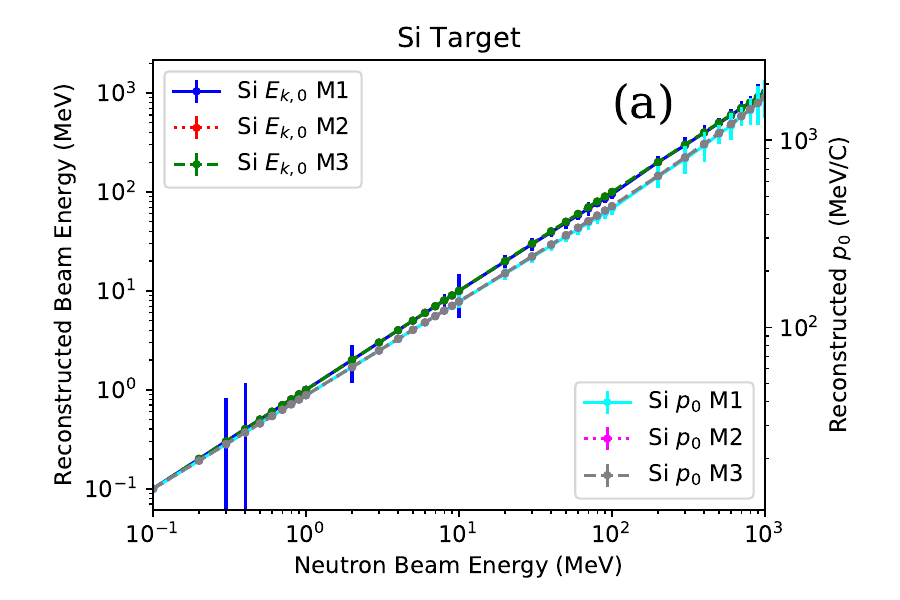}
    \includegraphics[width=0.49\textwidth]{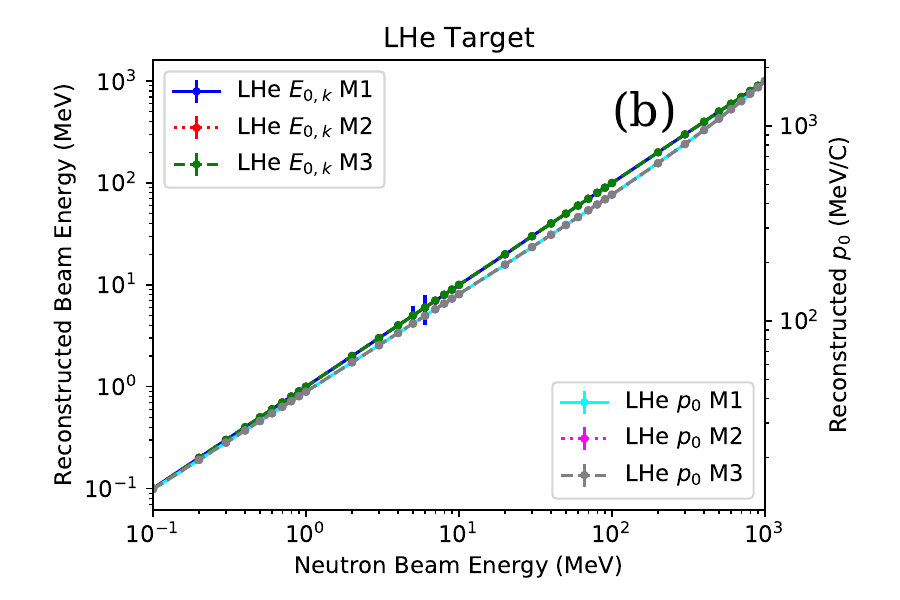}
\caption{The reconstructed initial momentum $p_0$ (green solid and magenta dotted lines) and beam energy $E_{k,0}$ (blue solid and red dotted lines) of neutrons for M1 (solid) and M2 (dotted). (\textbf{a}) Silicon target. (\textbf{b}) Liquid helium target. }
  \label{fig:Ep}
\end{figure}

Figure~\ref{fig:Ep} shows the reconstructed beam momentum $p_0$ using Equations~\eqref{eq:p0_M1} and~\eqref{eq:p0_M2} and the reconstructed beam kinetic energy $E_{k,0}$, where
\begin{align}
    E_{k,0} = E_{0}-m_n c^2 = \sqrt{(m_n c^2)^2+(p_0 c)^2}-m_n c^2
\end{align}
using the M1 and M2. The reconstructed beam kinetic energy is exactly equal to the input beam energy with a small error bar. This demonstrates that M1 needs A, B, C, $\vec{Q}_1$, and $\vec{Q}_2$,  M2 needs A, B, $\vec{Q}_1$, and $\tau_1$, and M3 needs A, B,  $E_{k,Q_{1}}$, and $\tau_1$ to derive the initial \mbox{beam momentum. }

\section{Ion Range}
\label{sec:ionrange}
\begin{figure}[h]
    \includegraphics[width=0.5\textwidth]{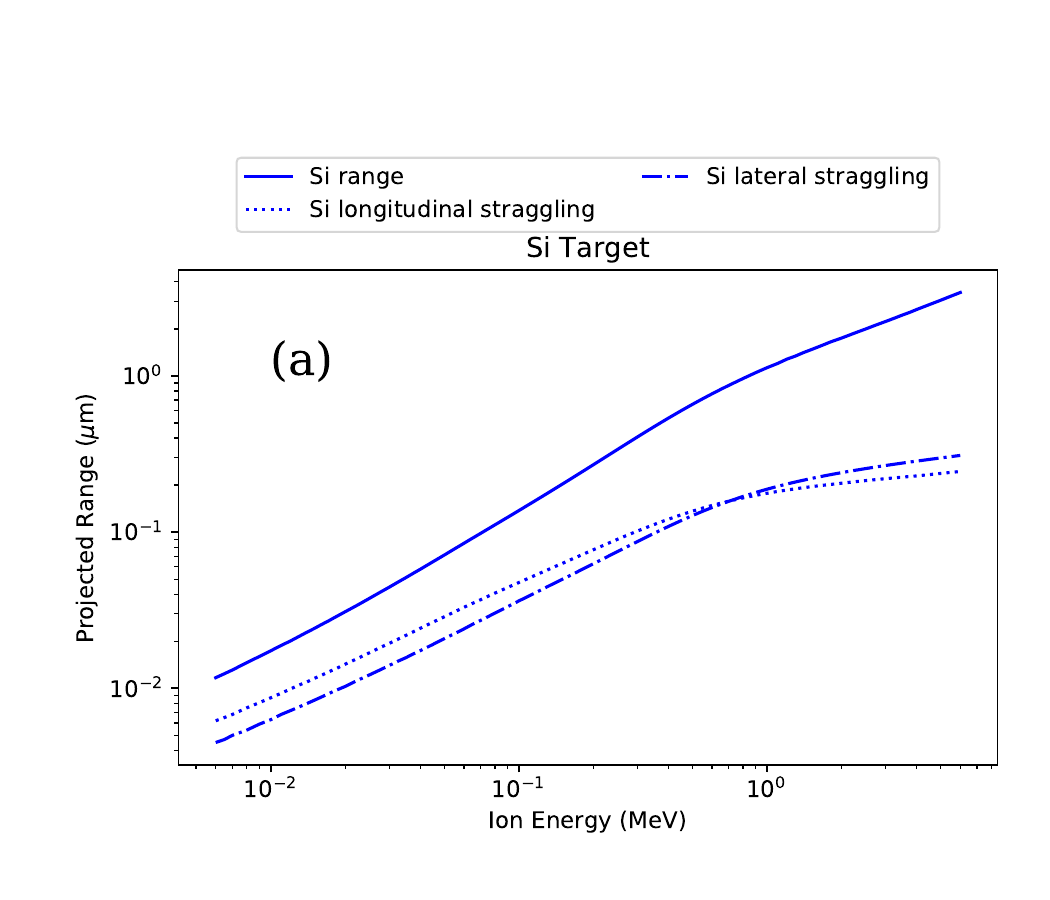}
    \includegraphics[width=0.48\textwidth]{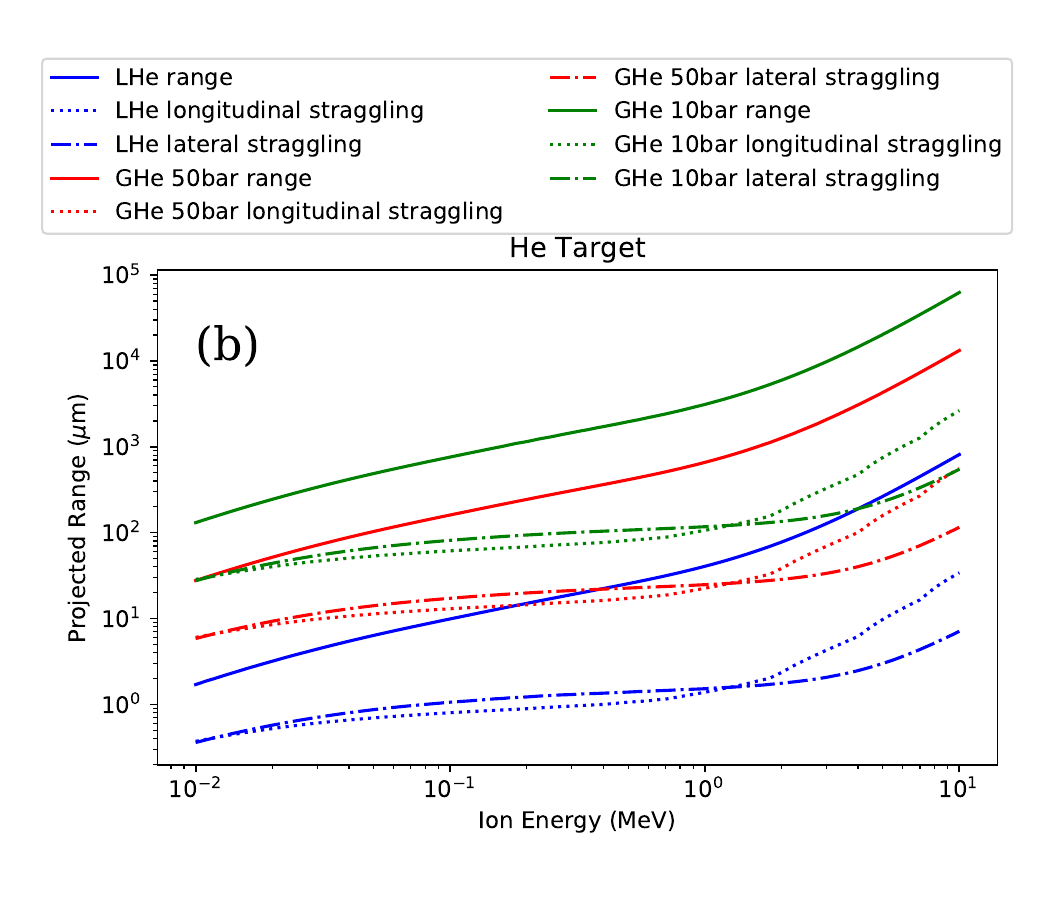}
\caption{The ion range table using SRIM. (\textbf{a}) Silicon ion in silicon target. (\textbf{b}) Helium ion in $^4$He target of liquid helium, gas helium of 50 bar, and gas helium of 10 bar. }
  \label{fig:ionrange}
\end{figure}

Figure~\ref{fig:AC_AB} shows the mean free path scale $\lambda$ of the neutron elastic collision around 5 cm in the silicon and 50 cm in the $^4$He. We also need to understand the ion travel range $L$ in order to estimate the position resolution of the detector. As shown in Figure~\ref{fig:Q1Q2_ionerg}, the silicon ion energy is about from 10 keV to 200 keV for the neutron energy from 100 keV to 10 MeV and above, and the $^4$He ion energy is about from 10 keV to 10 MeV for the neutron energy from 100 keV to 100 MeV and above. However, there is no ion information saved in PTRAC. Here, we apply the software of The Stopping and Range of Ions in Matter (SRIM)~\cite{Ziegler:1985} to calculate the ion range for given energies, as shown in Figure~\ref{fig:ionrange}. Silicon ions can only travel around 0.01 $\mu$m for the ion energy around 10 keV and 0.3 $\mu$m for the ion energy around 200 keV. $^4$He ions with energy 1 MeV can travel 40 $\mu$m in liquid helium, 700 $\mu$m in gas helium of 50 bar, and 3 mm in gas helium of 10 bar. Therefore, the scale of $L$ is about from 10 nm to 10 mm, depending on the targets and the neutron energy. Figure~\ref{fig:Q1Q2_ionerg} also shows the ion longitudinal straggling range $\delta L_{sl}$ and the ion lateral straggling range $\delta L_{st}$, which are about from nm to mm, about 1 order smaller than $L$. The detector position resolution must match at least the scale of $\delta L_{sl}$ and $\delta L_{st}$.  Since the $^4$He position resolution is about 100 $\mu$m, Figure~\ref{fig:ionrange} shows that gaseous $^4$He time projection chambers (TPC) will be ideal for ion tracking as well as tracking fast neutrons, which has been discussed in Ref.~\cite{WANG:2013145}. However, TPC could be expensive to construct and implement. On the other hand, silicon pixel detectors are promising for ion tracking with high-energy neutrons and, in particular, if micron-size pixels and voxels, which would be available in 1 cm$^3$ formats~\cite{wang:2021}. Additionally, the ion energy range with the silicon target is above keV in Figure~\ref{fig:ionrange} and the timescale is above 0.1 nanosecond in Figure~\ref{fig:TimeAB}. The energy and the time are about the proper scale for silicon pixel detectors. Therefore, the silicon pixel detectors can be used to measure neutrons with M3 as well. When designing a neutron tracking detector, this should be considered in designs. Different solid-state detectors are also investigated in Refs. ~\cite{Weinfurther:2018, Musumarra:2021}.


\section{Conclusions}
In this paper, we use MCNP to simulate the successive elastic neutron scattering in the natural silicon and $^4$He targets and demonstrate the robustness of tracking fast neutrons. The efficiency can be as large as 10\% for the neutron energy less than 10 MeV, but much smaller for high-energy neutrons for the moderate-size cube detectors less than 1 m$^3$. The ion travel range around mm is suitable for gaseous TPC detectors while the ion travel range around $\mu$m is only good for the silicon pixel detectors with high-energy neutrons. Gaseous TPC will be ideal for the fast neutron tracking; however, it could be expensive to construct and implement. Depending on the application of the detector, solid-state pixelated detectors can be useful for the fast neutron tracking, especially if high-position resolution around 1 micron can be achievable. At least three different methods to measure the initial neutron energy and direction can be applied for two different kinds of detectors, depending on the constraints of the detectors and the experimental conditions. A possible silicon pixel detector for the neutron tracking measurement will be constructed and investigated in the future.

\section*{acknowledgments}{The authors acknowledge the support from Dr. Bob Reinovsky of the Advanced Diagnostics or C3 program in Los Alamos National Laboratory. This work was supported in part by the U.S. Department of Energy through the Los Alamos National Laboratory (The Advanced Diagnostics or C3 program under the leadership of Dr. Bob Reinovsky). Los Alamos National Laboratory is operated by Triad National Security,LLC, for the National Nuclear Security Administration of U.S. Department of Energy (Contract No.89233218CNA000001).}

\bibliographystyle{elsarticle-num} 
 \bibliography{main}





\end{document}